\documentclass[12pt,preprint]{aastex}
\newcommand{\fouru}   {{4U~0142+61}} 
\newcommand{\chan}   {{\sl Chandra}}

\newcommand{\asca}   {{\sl ASCA}}

\newcommand{\cxo}    {{\sl Chandra X-ray Observatory}} 
\newcommand{\nh}     {$N_{\rm H}$} 
 
\newcommand{\kt}     {$kT_{\rm BB}$} 
 
\begin{document}
\title{\chan\ Observations of the Anomalous \\ X-ray Pulsar \fouru } 
\author{ 
Sandeep~K.~Patel\altaffilmark{1}, 
Chryssa~Kouveliotou\altaffilmark{2,3}, 
Peter~M.~Woods\altaffilmark{2}, 
Allyn~F.~Tennant\altaffilmark{3}, 
Martin~C.~Weisskopf\altaffilmark{3}, 
Mark~H.~Finger\altaffilmark{2}, 
Colleen~Wilson-Hodge\altaffilmark{3}, 
Ersin~{G\"o\u{g}\"u\c{s}}\altaffilmark{4},
Michiel~van~der~Klis\altaffilmark{5}, 
Tomaso~Belloni\altaffilmark{6} 
} 
\altaffiltext{1} {National Academy of Sciences-NRC/NSSTC, SD-50, 
Huntsville, AL 35805} 
\altaffiltext{2} {Universities Space Research Association/NSSTC, SD-50, 
Huntsville, AL 35805} 
\altaffiltext{3} {NASA Marshall Space Flight Center, Huntsville, SD-50, 
AL 35812} 
\altaffiltext{4} {Department of Physics, University of Alabama in 
Huntsville/NSSTC-SD50, Huntsville, AL 35805} 
\altaffiltext{5} {Astronomical Institute "Anton Pannekoek," University 
of Amsterdam, and Center for High Energy Astrophysics, Kruislaan 403, 
1098 SJ Amsterdam, Netherlands} 
\altaffiltext{6} {Osservatorio Astronomico di Brera, Via Bianchi 46, 
23807 Merate (Lc), Italy} 

\begin{abstract} 
We present X-ray imaging, timing, and phase resolved spectroscopy of 
the anomalous X-ray pulsar \fouru\ using the \cxo\ (CXO). The spectrum is well 
described by a power law plus blackbody model with $\Gamma = 3.35(2)$, 
\kt$=0.458(3)$~keV, and \nh$=0.91(2)~\times~10^{22}~{\rm cm}^{-2}$; 
we find no significant evidence for spectral features ($0.5-7.0$ keV).
Time resolved X-ray spectroscopy shows evidence for evolution in phase in either $\Gamma$, or \kt\, or some combination thereof as a function of pulse phase. We derive a precise X-ray position for the source and determine its spin 
period, $P=8.68866(30)$~s. We have detected emission beyond $4\arcsec$ from the central source and extending beyond $100\arcsec$, likely due to dust scattering in the interstellar medium. 

\end{abstract} 
\section{Introduction} 
\label{sec:intro} 
\fouru\ was discovered in 1978 with {\em Uhuru} \citep{Forman78}, but
only in 1984 were 8.7-s pulsations detected in its persistent X-ray
flux with {\em EXOSAT} \citep{Israel94}.  
Pulse frequencies were sparsely measured until
the last few years.  The frequency values are
consistent with a constant spin-down rate of $\dot \nu = (-3.0\pm 0.1)
\times 10^{-14}$ Hz s$^{-1}$, slightly larger than the $\dot\nu =
(-2.598 \pm 0.002) \times 10^{-14}$ Hz s$^{-1}$ recently measured with
{\em RXTE} phase connected data \citep{Gavriil01}. Despite detailed
searches \citep{Israel94,Wilson99}, no evidence for orbital Doppler
shifts has been found in the pulse frequency measurements. Upper
limits of $a_{\rm X} \sin i \lesssim 0.26$ lt-s for orbital periods in
the range 70 s to 2.5 days \citep{Wilson99} place stringent limits on
allowed orbital inclinations and companion masses for normal or helium
main sequence companions. 

{\em ASCA} observations in 1994 \citep{White96} and in 1998
\citep{Paul00}, both determined the source energy spectrum as an
absorbed power law (photon index $\sim 3.7$) with a blackbody ($kT
\sim 0.4$ keV) component. The {\em ASCA} GIS data also revealed very 
stable (in time) pulse profiles and intensities. The pulse profiles, 
however, evolved with energy from double-peaked (0.5-4.0 keV), to 
single peaked (4.0-8.0 keV) \citep{White96}. Later spectral observations 
with {\em BeppoSAX} in 1997-1998 \citep{Israel99} were consistent with 
the {\em ASCA} spectral results, however it did not show significant 
pulse phase dependence.  Using RXTE observations, \cite{Gavriil01} reported 
pulse profile evolution with energy, evolving from 
double-peaked (2.0$-$4.0 keV) to single peaked (6.0$-$8.0 keV). 
\cite{Juett02} used the High Energy Transmission Grating Spectrometer (HETG) on the CXO to observe the source in 2001.  They find no evidence for emission or absorption lines in the spectrum which is well fitted by an absorbed powerlaw$+$blackbody with $\Gamma=3.3(4)$ and $kT=0.418(13)$ keV. Their analysis used only phase averaged spectra and provided stringent constraints on spectral features.

Recent optical observations \citep{Hulleman00} have revealed an
object with peculiar optical colors inside the {\em Einstein} 
error circle \citep{White87}. The optical observations imply
an extinction of $2.7 \lesssim A_{V} \lesssim 5.0$ and a distance of
$d \gtrsim 2.7$ kpc. The object is too faint to be a large accretion
disk. Further observations of the optical counterpart revealed 
pulsations with a 30\% pulse fraction \citep{Kern01}. The large pulse
fraction also rules out X-ray reprocessing (in a disk) as the source
of the optical pulsations. A deep radio search showed no evidence for
a SNR, extended structure, or even a point source at the position of
\fouru\ \citep{Gaensler01}. 

We present here an analysis and discussion of our \chan\ observation
of the source. In \S \ref{sec:obsresults} we describe the x-ray
observations, and present spatial, timing and spectral results of
\fouru.  We also include spatially resolved spectroscopy of the
extended emission around \fouru, presumably due to dust scattering.  A
precise x-ray position is derived and compared with the optical
identification from \cite{Hulleman00}.  
~

\section{Observations and Results} 
\label{sec:obsresults} 

We observed \fouru\ on 21 May 2000 with the \chan\ Advanced CCD
Imaging Spectrometer (ACIS).  Data were collected sequentially in two
different observing modes: timed exposure (TE) mode  and continuous
clocking (CC) mode. Data obtained in TE mode allow for two-dimensional
imaging but not for accurate time averaged or pulse-phased
spectroscopy due to pulse pile up and the low (3.24~s)
time resolution of this data type.  In CC mode the amount of pile up
is negligible and one can exploit the 2.85~ms time resolution though
the image is one-dimensional. The source was positioned on the nominal
target position of ACIS-S3, a back illuminated CCD on the
spectroscopic array (ACIS-S) with good charge transfer efficiency and
spectral resolution. 

Standard processing of the data was performed by the \chan\ X-ray
Center to Level 1 (CXC processing software version R4CU5UPD13.3).  
The \chan\ calabration database (CALDB2.9) was
utilized and the data were further processed using the CIAO (v2.2.1)
tool {\sl acis\_process\_events}. The data were filtered to exclude
events with \asca\ grades 1, 5, and 7, hot pixels, bad columns, and
events on CCD node boundaries.  The S3 light curve was inspected in a
region offset from the AXP to identify times of high background rate.
We binned the data in 200~s intervals and found no times when the
background exceeded a $3\sigma$ threshold about the mean, and so all the
data were deemed useful. The resulting observing times are 7744~s and
5945~s for TE and CC mode data, respectively. Uncertainties are 
reported at the $68\%$ confidence level.

\subsection{Timing Results} 
\label{sec:axptim} 

High resolution timing with ACIS is only possible using CC-mode data.
The CC mode event times denote when the event was read out of the
frame store, not when it was detected. We corrected for this effect by
assuming that all photons were originally detected at the nominal
target position.  We removed the variable time delay due to spacecraft
dither and telescope flexure. The event arrival times were then
corrected to the solar system barycenter using the JPL planetary
ephemeris DE-200. 

The data were divided into ten intervals, each of $\sim600$~s, and
individual pulse profiles were created using a coarse ephemeris 
derived from an epoch fold search.  These ten profiles
were compared to the pulse profile derived using all the data and the
relative phases were fit with a linear function.  The resulting pulse
period of $8.68866\pm0.00030$~s is referred to epoch MJD 551685.78.
This period is consistent with the more precise spin history of the
source determined with long term RXTE monitoring \citep{Gavriil01}.

\begin{figure*}[r] 
\epsscale{1.0} 
\plotone{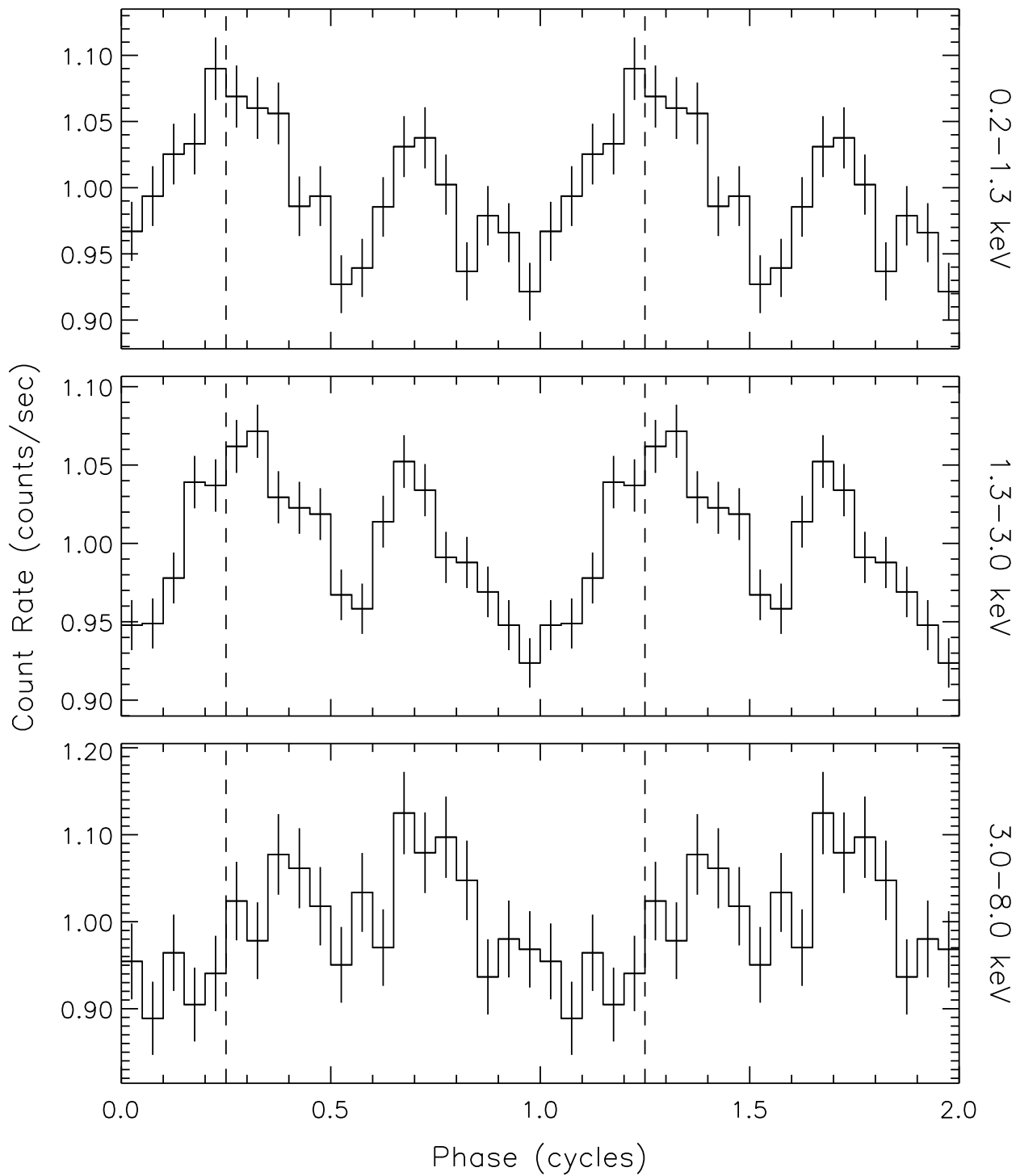}
\caption{ The pulse profile of \fouru\ in three energy ranges. The
RMS and peak to peak (PTP) values for each energy range are {\sl Top}:
$0.2-1.3$~keV, RMS $=4.6\pm0.5 \%$ , PTP $=8.4\pm1.6 \%$,  {\sl Middle}: $1.3-3.0$~keV, RMS $=4.1\pm0.4 \%$ , PTP $=7.4\pm1.2 \%$,  and {\sl Bottom}: $3.0-8.0$~keV, RMS $=5.6\pm1.0 \%$ , PTP $=11.7\pm3.2 \%$.} 
\label{fig:pulsefraction} 
\end{figure*}

Hereafter we use the entire dataset for our timing
analysis.  The pulse profile (0.5$-$7.0 keV) is shown in
Figure~\ref{fig:foldevents}a; the peak-to-peak pulse fraction
($PTP \equiv \frac{F_{\rm max}-F_{\rm min}}{F_{\rm max}+F_{\rm min}}$)
is $6.7\pm 0.9\%$.  Here, $F_{\rm max}$ and $F_{\rm min}$ are the
maximum and minimum count rates. We present in
Figure~\ref{fig:pulsefraction} the variation in both the $rms$ and
peak-to-peak pulse fraction of the profile with energy. We find no
significant variation of these values in three
energy channels ($0.1-1.3$, $1.3-3.0$, and $3.0-8.0$ keV). However, we
see evidence for evolution in the profile shape with energy (see \S 2.4 
for quantitative details) with the first pulse being the softest. 

\subsection{AXP Position} 
\label{sec:position}

We use only data collected in TE mode to derive the AXP position. 
For \fouru, which is rather bright ($\sim 21$ ACIS counts/s between
$0.5-7.0$ keV), the data suffer from pile-up of
photons in the image core, as a result of which the source image looks
ring-shaped with a hole in the center.  We modeled the data using a
Gaussian multiplied by a hyperbolic tangent in radius, scaled to
approach zero at $r=0.0$ \citep{Hulleman01}.  
The resulting best-fit centroids
correspond, using the \chan\ aspect solution, to a J2000
position on the sky of $\alpha=01^{\rm h}46^{\rm m}22\fs42$,
$\delta=+61\arcdeg45\arcmin02\farcs8$.  The uncertainty is limited by
systematic effects, to a circle with $\sim\!0.7\arcsec$ radius
\citep{aldckc+00}.   This position is consistent
with the location of the optical counterpart found by \cite{Hulleman00},
$\alpha=01^{\rm h}46^{\rm m}22\fs41$,
$\delta=+61\arcdeg45\arcmin03\farcs2$.  Unfortunately, only one other
(faint) X-ray source is visible in the TE observation. We were,
therefore, unable to reduce the \fouru\ positional error by using
astrometry to obtain a boresight correction.

\subsection{Extended Emission Search} 
\label{sec:extended} 

While searching for extended emission (within a few
arcseconds) one must account for numerous contributions to
the observed flux. These contributions include not only the bright
source itself and the instrumental background, but diffuse emission 
from the galactic plane, a dust scattering halo, and the potential
contribution of an X-ray nebula.

To avoid issues related to the effects of pile up in the TE mode data,
we first utilized the one dimensional image from the CC mode. We
generated a time-integrated image ($0.5-7.0$ keV) minus any point
sources other than \fouru. We then subtracted an average count rate
intended to remove instrumental and diffuse cosmic X-ray background 
\citep[][and references therein]{markevitch2001}.

Next, we constructed what will be referred to as the ``pulsed'' image.
We took the observed pulse profile, normalized it to a mean of zero
and convolved it with the event list of S3.  Each count recorded on S3
was assigned a phase. Events with a phase near pulse maximum,
regardless of their position on the chip, received a positive weight
and likewise, those near pulse minimum, a negative weight.  In doing
so, we remove all emission components in our image that do not vary in
phase (i.e., everything except the central pulsar).  Note that the
time delay for photons scattered by the interstellar dust is on
average minutes, much longer than the pulsar period, and therefore will
not contaminate the pulsed image. 

\begin{figure*}[r] 
\epsscale{1.0} 
\plotone{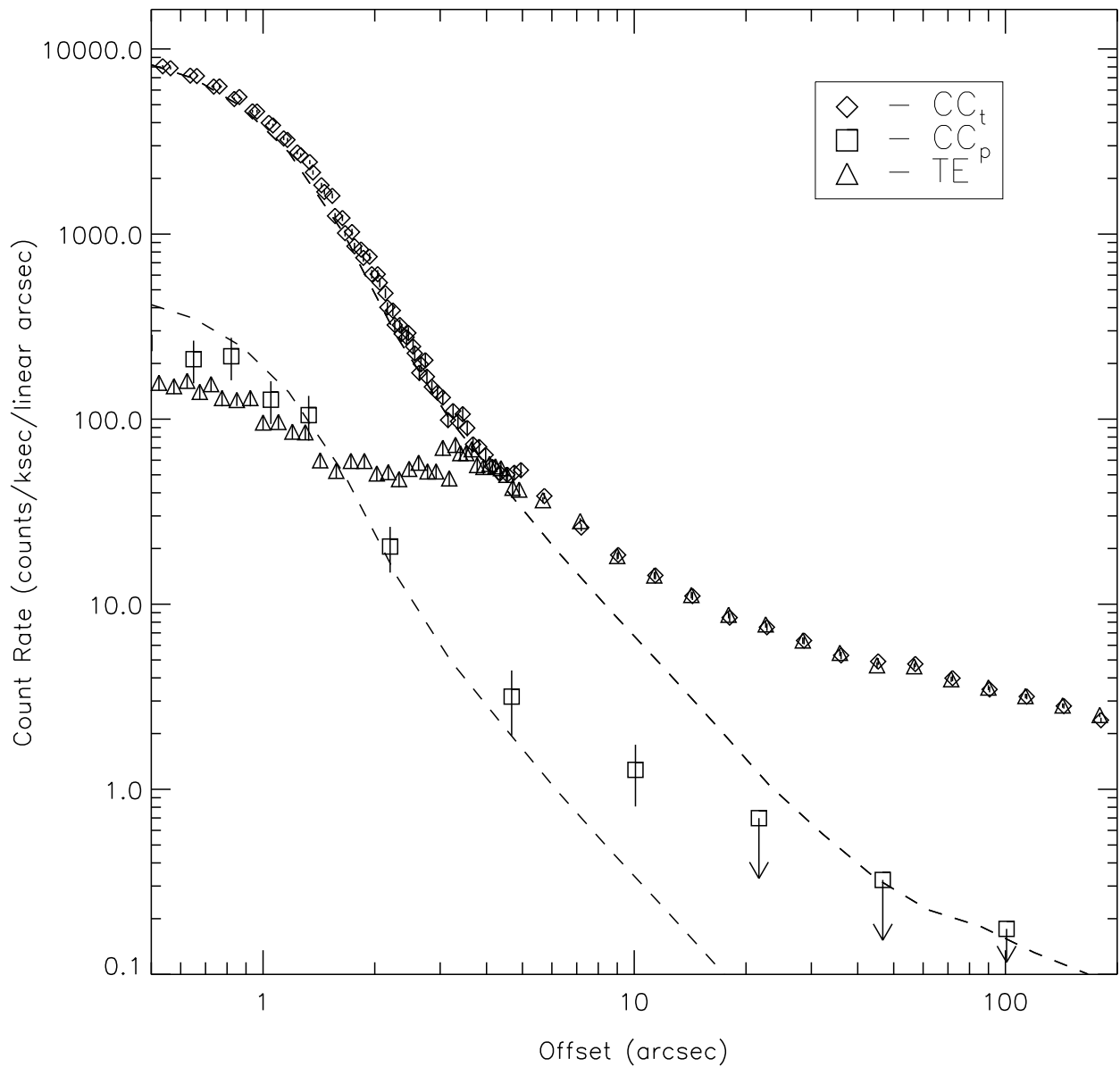}
\caption{Radial surface brightness profiles of the total (CC$_{\rm t}$)
and pulsed (CC$_{\rm p}$) emission (0.5$-$7.0 keV).  The dashed lines
are the simulated \chan\ point spread function. Downward pointing arrows
denote 2$\sigma$ upper limits to the count rate. The triangles are  TE
mode data. Note the agreement between the total CC profile and the TE
profile at radii $\gtrsim4\arcsec$.} 
\label{fig:radprof} 
\end{figure*}

To improve statistics, we then folded the 1-D time-integrated
(CC$_{\rm t}$) and pulsed (CC$_{\rm p}$) images about the common
centroid and accumulated ``quasi-radial'' profiles.  These profiles
are shown in Figure~\ref{fig:radprof}.  We compare the pulsed radial profile
with the MARX \citep[v3.01]{wise2000} derived point spread function (PSF) and calculate $\chi^2/\nu = 15.2/8$.  We conclude that the pulsed radial profile
 is consistent with the MARX PSF. 

Next, we collapsed the TE mode observation to mimic the 1-D CC image
(Figure~\ref{fig:radprof}, TE). We find that the TE profile completely
overlaps the CC$_{\rm t}$ profile beyond $\sim 4\arcsec$, marking the
radius beyond which the effects of pile-up are negligible. 
We conclude that the majority of the \fouru\ X-ray flux is contained
within a radius of $\sim 4\arcsec$. 

In both profiles we find excess emission beyond $\sim 4\arcsec$ which is
likely due to a dust-scattering halo and potentially an X-ray plerion.
Azimuthally averaged radial profiles generated in different energy 
bands and spectra extracted from S3 in 3 concentric annuli centered 
on the pulsar position supports the dust scattering halo hypothesis. 
Detailed modelling of the spatial and spectral 
distribution of the dust scattered emission from a sample of X-ray 
bright sources, including \fouru, is underway and will be presented 
elsewhere. 

\subsection{Spectral Analysis} 
\label{sec:axpspec} 

We use CC mode data to obtain an X-ray spectrum as these are not impacted by pile up. As no response calibration is currently available for \chan\ CC mode data, here we assume that the CC and TE spectral responses are identical. To test this assumption, we compared spectra extracted from both CC and TE mode data. The extraction regions were selected from identical parts of the sky, more than $5\arcsec$ from \fouru, and close enough to the AXP to contain flux from the dust scattering halo.  Background spectra were extracted from a region on S3 offset from this source region.  All spectra and response files for our analysis were generated using the CIAO (v2.2.1) tools {\sl dmextract}, {\sl mkrmf}, and {\sl mkarf}, and CALDB(v2.9). The extraction was performed in pulse invariant (PI) space ({\em{i.e.}}, after the instrument gains were applied). We fit both CC and TE mode spectra with a TE mode response and an absorbed power law model; we found acceptable fits and reasonable agreement in the derived model parameters (CC: $N_{\rm H} = 1.1\pm0.1$, $\Gamma=3.2\pm0.2$, $\chi^2/\nu= 82.3/80$, and TE: $N_{\rm H} = 1.1\pm0.1$, $\Gamma=3.1\pm0.2$, $\chi^2/\nu= 84.8/103$).  As a seperate check, we subtract the CC mode spectra from the TE mode data and fit the difference to a constant.  This fit resulted in a $\chi^2/\nu=109.4/106$ and was consistent with zero.  Based on these tests, we find it valid to adopt the TE mode response in our spectral analysis of the CC mode data. 

We define the AXP source region by selecting an interval $\pm 8$ pixels ($\pm
\sim 4\arcsec$) (see also \S \ref{sec:extended}) around the peak flux
along the collapsed CC mode axis.  The background was determined using
two adjacent segments 16 pixels wide for a total area 2 times the
source area; the background flux contributes $\sim1.7\%$ of the total 
(0.5-7.0 keV) flux in the source region. 

\begin{figure*}[r] 
\epsscale{0.8} 
\plotone{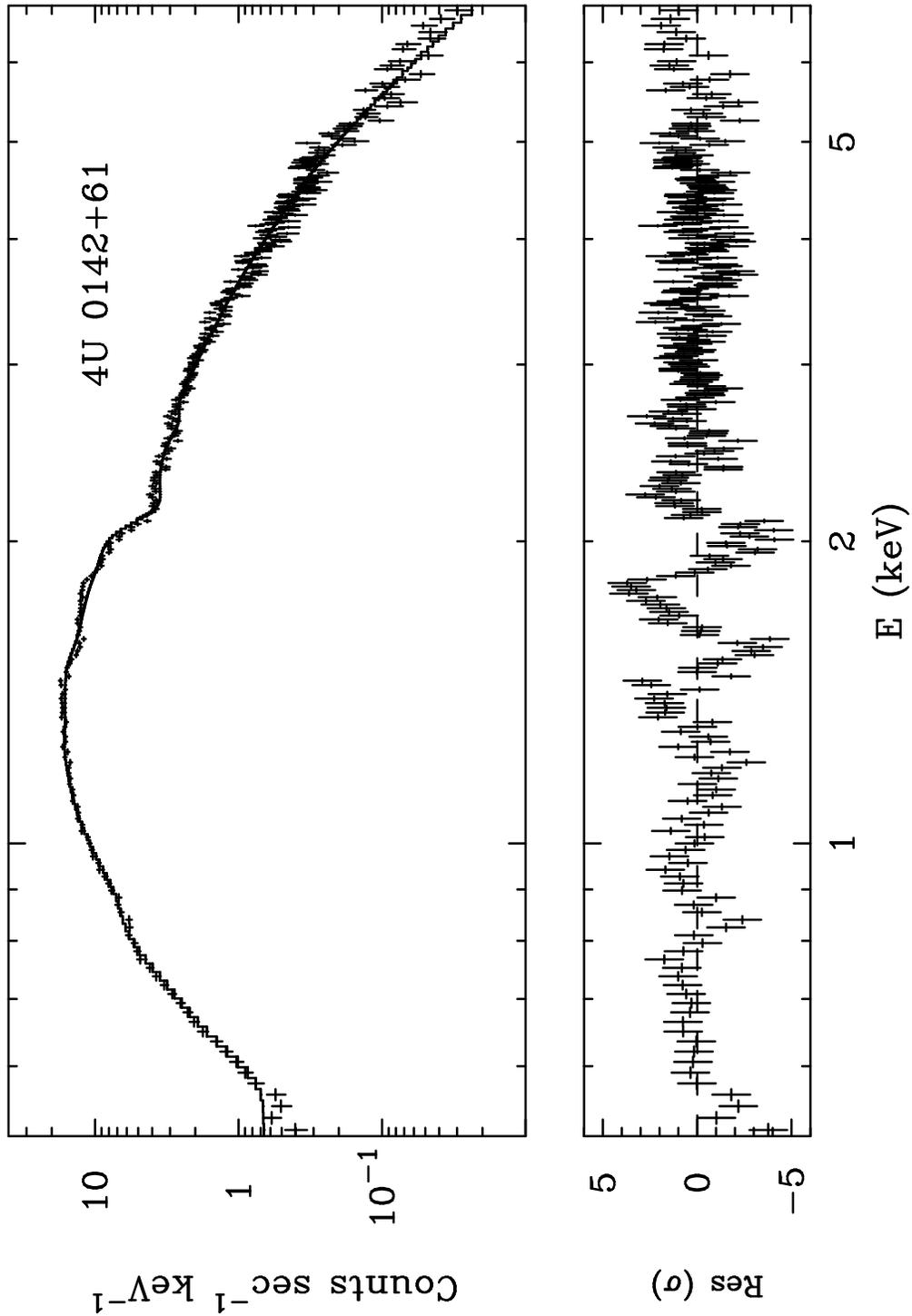}
\caption{The best fit phase averaged spectrum of \fouru.  The spectral
data and model (PL$+$BB) are shown in the top panel and its residuals
in units of $\sigma$ are shown in the bottom panel. The feature at
$\sim2.0$~keV is due to a small shift of the location of the Ir
absorption edge between the response and the data.}
\label{fig:avespec} 
\end{figure*}

The phase averaged spectrum of the source is shown in
Figure~\ref{fig:avespec}. We have fit several models to the spectrum
and we find that the best fit model is an absorbed power law +
blackbody function, which gives $\Gamma = 3.40\pm0.06$,
$kT_{\rm BB} =0.470\pm0.008$, with a hydrogen column
density, $N_{\rm H} = 0.93\pm0.02$.  These parameter values 
are consistent with the values derived from simultaneous fits to phase 
resolved spectra (this is discussed in more detail later in this section). 
All spectral fits were
limited to the $0.5-7.0$~keV band with XSPECv11.01 \citep{arnaud1996},
and use the photo-electric absorption coefficients of
\cite{Morrison83},  and abundances of \cite{anders82}.  

Our results for \nh\ and $\Gamma$ are consistent with the 
phase averaged results of
\cite{Juett02} derived from \chan\ HETG data.  However, the blackbody
temperature derived here differs slightly ($3.7 \sigma$) and is higher 
than the one derived from the HETG data. We do not view this difference 
as compelling since there are systematic effects that have been treated 
differently in the two analyses.  The blackbody temperature 
derived here is also higher than the one derived using \asca\ data 
\citep{Paul00}; the difference is likely due to contamination from 
the scattering halo within the large (3$^{\prime}$ HPD) point 
spread function of the \asca\ telescopes. 

We search for evolution of the spectrum with pulse phase by
systematically looking for significant variations in fitted spectral
parameters as a function of pulse phase.  Using the derived pulse 
period (\S \ref{sec:axptim}) we
construct a source spectrum for each of ten pulse phase bins.  
Each of the 10 resulting spectra were grouped into bins that 
contained at least 25 events. 
 
We constrain the \nh\ to be identical at each phase 
(hereafter defined as ``linked''), since we do not expect large variations 
in the absorption over the small angular size of the source 
or any significant intrinsic absorption.  

To confirm this assumption we perform and compare similar joint fits
with (a) \nh\ free to aquire a best fit value in all phase bins and
(b) \nh\ linked at each phase.  All other model parameters are free to
obtain their best fit values in each phase bin.  We then fit the data
from each of the ten pulse phase bins with an absorbed PL+BB model.
The former method provides similar values of \nh\ for all bins and
consistent with the linked value of $N_{\rm H} = 0.91\pm0.02$
derived with the latter method.  Furthermore, the former fit results in
a $\Delta \chi^2 = 6.1$ for 9 additional parameters; an $F-$statistic
probability of $78\%$ is calculated, indicating that allowing \nh\ to
be free does not statistically improve the fit.  

Keeping \nh\ linked thereafter, we initially we fit the data with the
power law normalization ($PL_{Norm}$), index ($\Gamma$), blackbody
normalization ($BB_{Norm}$), and \kt\ free (Table \ref{tab:1}, model
5); we found a $\chi^2/\nu= 1950.94/1779$.  The fit is
significantly improved ($\chi^2/\nu=1491.469/1429$) by ignoring the
$1.64-2.14$ keV band, which contains the iridium-edge structure in the
telescope response (2.04 keV) and the Si K fluorescence line (1.74
keV) and Si absorption edge (1.84 keV).  (Current gain calibration is
known to cause spurious features in high S/N spectra in this energy
range\footnote{http://asc.harvard.edu/udocs/}). One approach to dealing 
with this situation is to eliminate data from the analysis.  However, to 
avoid eliminating data, we have opted for an alternative method of analysis. 
We included all events with energy between 0.5-7.0 keV and
accounted for systematic errors by using the calculated reduced
$\chi^2$ ($\chi^2_{\nu}$ to determine our 1-$\sigma$ error bound.  If
$\chi^2_{\nu} \leq 1$ then the range of parameter space that
encompasses a $\Delta \chi^2 = \pm 1$ is adopted as the 1-$\sigma$
uncertainty; if $\chi^2_{\nu} > 1$ then the range of parameter space
that encompasses a $\Delta \chi^2 = \pm \chi^2_{\nu}$ is adopted as
the 1-$\sigma$ uncertainty. 

The above is the least restrictive constraint to the data and shows evidence
for evolution with phase in \kt, but not in $\Gamma$. To investigate
the evolution of spectral parameters, we perform a series of more
constraining fits to the data.  

We begin with an extremely restrictive fit where \kt\ and $\Gamma$ are
linked and the ratio of their normalizations is constant.    The
results of this fit are shown in Table \ref{tab:1} as model 1.  

We next assume (Table \ref{tab:1}, model 2) that the spectral shape is 
unchanged, but the contribution of each component varies.
Consequently, we fit all phase bins 
with $\Gamma$ and \kt\ linked, and allow the normalizations to vary. 
This fit results in a lower $\chi^2$ as compared to model 1, 
($\Delta \chi^2=16$, $\Delta \nu=10$) which only improves the fit at 
the $84\%$ confidence level.  Hence, for the
remaining fits we constrain the ratio of the normalizations to be linked. 

\begin{figure*}[r] 
\epsscale{1.0} 
\plotone{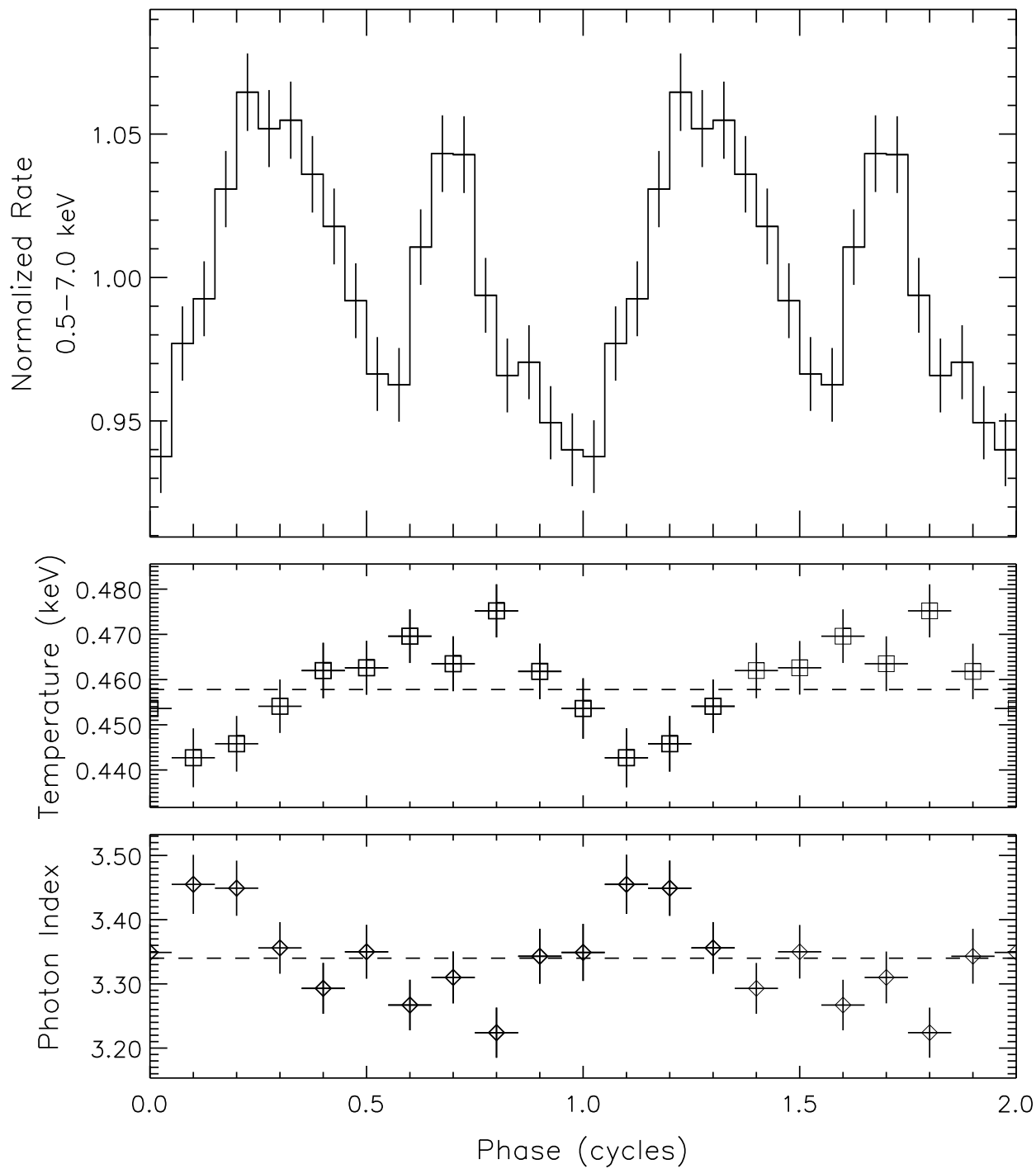}
\caption{Top panel (a): The pulse profile of \fouru\ ($0.5-7.0$~keV). 
Middle panel (b): Variation of the blackbody temperature from 
fitting each phase bin with a PL+BB model (Model 1). The phase 
averaged value is denoted by the dashed line (Model 5). 
Bottom panel(c) : Variation of the PL index (Model 1) and phase 
averaged value (Model 5: dashed line).} 

\label{fig:foldevents} 
\end{figure*}

Next, we keep $\Gamma$ linked and allow \kt\ and the normalizations to
vary (Table \ref{tab:1}, model 3).  The decrease in $\chi^2$ here is
significant ($\Delta \chi^2 = 34.2$, $\Delta \nu = 9$) indicating that
allowing \kt\ to vary statistically improves the fit. The evolution of
blackbody temperature is shown as a function of pulse phase in
Figure~\ref{fig:foldevents}b. The spread in \kt\ values is $0.032$ keV.  

The fits were repeated with \kt\ linked and with varying $\Gamma$ and the 
normalizations (Table \ref{tab:1}, model 4).   The fit results in a
$\Delta \chi^2=35.1$ ($\Delta \nu=9$) which indicate that allowing
$\Gamma$ to vary, also statistically improves the fit. The evolution of the 
power law index is shown as a function of pulse phase in 
Figure~\ref{fig:foldevents}c. The range of $\Gamma$ variations here is 
0.24.  Models 3 and 4 indicate that there is evidence for evolution in phase 
in either $\Gamma$, or \kt\, or some combination thereof, as indicated by 
Table \ref{tab:1}, model 5.   The average 2$-$10 keV unabsorbed flux and X-ray luminosity, calculated using model 5, are $8.3~\times~10^{-11}$~ergs~s$^{-1}$~cm$^{-2}$ and $0.99~d_{kpc}^2\times~10^{34}$~ergs~s$^{-1}$, respectively. The PL component contributes $61\%$ to the 2$-$10 keV unabsorbed flux. 

Finally, we have searched for spectral lines, but find no significant
candidates. Features that appear below $\sim 1$~keV and in the Ir edge, are potentially due to uncertainty in the spectral response. To set upper limits, we examined two small deviations at 1.0 and 4.9~keV. The addition of a line, modeled by a Gaussian with an intrinsic width of 40 eV at 4.9 keV to the PL+BB model (Table \ref{tab:1}, model 5), results in a $\Delta \chi^2 = 6.4$ for 3 additional parameters; an $F-$statistic probability of $95\%$ is calculated, the $90\%$ confidence upper limit on the line flux is $4.3 \times 10^{-13}$~ergs~cm$^{-2}$~s$^{-1}$.   Similarly, including a ``line'' at 1.0 keV results in a $\Delta \chi^2 = 4.4$ and and $F-$statistic probability of $86\%$.  Our results are consistent with the results of \cite{Juett02} derived from \chan\ HETG data.

\section{Discussion} 
\label{sec:discuss} 

We have observed the anomalous X-ray pulsar, \fouru\ with \chan\ and 
performed detailed phase resolved spectroscopy of the source. We find
that there is significant spectral evolution during the overall
pulse.  The phase averaged spectrum can be best fit with a two
component model (PL+BB), consistent with earlier results
(\cite{Juett02,White96,Israel99,Paul00}). We confirm the source location of
\cite{Juett02}, which also coincides with the optical counterpart of
\cite{Hulleman00}. We measure a pulse period consistent with the 
measurements of \cite{Gavriil01} and confirm limits to the amplitude of 
possible binary motion in a subset of period ranges previously investigated 
by \citep{Wilson99}.

The superb \chan\ spatial resolution allows us to determine that  
there is emission extending up to $100\arcsec$ from the pulsar, likely 
due to dust scattering.
We are unable to determine evidence for excess emission within a few 
arcseconds of the source (i.e., evidence for a plerion), due to 
contamination from the X-ray bright AXP and the associated dust 
scattering halo.   

\acknowledgments
 
We thank H. Marshall for allowing us to compare our results with their
observations prior to their publication, and for many insightful
comments. This work was supported by grants MX-0101 and GO0-1018X
(C.K.), NAG5-9350  (P.W.), GO0-1018X (S.P.).


\begin{deluxetable}{llllcc}
\footnotesize
\tablecaption{Phase Resolved Spectral Fits for \fouru
\label{tab:1}}
\tablewidth{450pt}
\tablehead{
\colhead{} & \colhead{Model Settings\tablenotemark{a}} &\colhead{} &
\colhead{} & \colhead{} \\
\colhead{{\rm Model ID}} &  \colhead{PL$_{\rm Norm} \propto$ BB$_{Norm}$} & \colhead{$kT_{\rm BB}$ Linked} & \colhead{$\Gamma$ Linked}         & \colhead{$\chi^2 / \nu$}  & \colhead{$F-$Test Probability\tablenotemark{b}} 
} 
\startdata
1 & Y & $0.458^{+0.003}_{-0.003}$  & $3.35^{+0.02}_{-0.02}$  & $2009.428 / 1807$  & {...} \\
2 & N         & $0.465^{+0.003}_{-0.008}$  & $3.39^{+0.01}_{-0.01}$  & $1993.472 / 1797$  & 0.843 \\ 
3 & Y         & N                          & $3.35^{+0.02}_{-0.02}$  & $1975.203 / 1798$  & 0.9997\\
4 & Y         & $0.458^{+0.003}_{-0.003}$  & N                       & $1974.285 / 1798$  & 0.9998\\

5 & Y         & N                          & N                       & $1963.639 / 1789$  & 0.9986 (0.624,~0.691)\tablenotemark{c}
\enddata
\tablenotetext{a}{Model normalizations for each phase bin are free
(unconstrained) to obtain its best fit value while $N_{\rm H}$ is
linked (constrained to be identical at each phase) for all model
variations. If the index or temperature are linked, then the best fit
values are listed with $1~\sigma$ uncertainties, else an ``N'' is
listed indicating the parameters are not linked. The Temperature is
listed in units of keV.}

\tablenotetext{b}{Based on the $F-$ statistic, we claim that the given
model is significantly better at the listed confidence levels.  The
probabilities are determined with respect to Model 1.}

\tablenotetext{c}{The numbers in parentheses are the probabilities
that Model 5 is significantly better than Models 3 and 4,
respectively.}  

\end{deluxetable}

\begin{thebibliography}{22}
\expandafter\ifx\csname natexlab\endcsname\relax\def\natexlab#1{#1}\fi

\bibitem[{{Aldcroft} {et~al.}(2000){Aldcroft}, {Karovska}, 
{Cresitello-Ditmar},  {Cameron}, \& {Markevitch}}]{aldckc+00} 
{Aldcroft}, T.~L., {Karovska}, M., {Cresitello-Ditmar}, M.~L., 
{Cameron},  R.~A., \& {Markevitch}, M.~L. 2000, \procspie, 4012, 650

\bibitem[Anders \& Ebihara(1982)]{anders82} Anders, E.~\& 
Ebihara, M.\ 1982, \gca, 46, 2363 

\bibitem[{{Arnaud}(1996)}]{arnaud1996}
{Arnaud}, K. 1996, in ASP Conf. Ser. 101: Astronomical Data Analysis
Software
  and Systems V, G. H. Jacoby \& J. Barnes (eds), 5, 17

\bibitem[Bildsten et al.(1997)]{Bildsten97} Bildsten, L.~et al.\ 1997,
\apjs, 113, 367

\bibitem[Forman et al.(1978)]{Forman78} Forman, W., Jones, C., Cominsky,
L., Julien, P., Murray, S., Peters, G., Tananbaum, H., \& Giacconi, R.\
1978, \apjs, 38, 357

\bibitem[Gaensler {et~al.}(2001)]{Gaensler01}
Gaensler, B.~M., Slane, P.~O., Gotthelf, E.~V., \& Vasisht, G.\ 2001,
\apj, 559, 963

\bibitem[Gavriil \& Kaspi(2001)]{Gavriil01} Gavriil, F.~P.~\& Kaspi,
V.~M.\ 2001, \apj accepted 

\bibitem[Ghosh, Angelini, \& White(1997)]{Ghosh97} Ghosh, P., Angelini,
L., \& White, N.~E.\ 1997, \apj, 478, 713

\bibitem[Hulleman, van Kerkwijk, \& Kulkarni(2000)]{Hulleman00}
Hulleman, F., van Kerkwijk, M.~H., \& Kulkarni, S.~R.\ 2000, \nat, 408,
689

\bibitem[Hulleman et al.(2001)]{Hulleman01} Hulleman, F., Tennant, 
A.~F., van Kerkwijk, M.~H., Kulkarni, S.~R., Kouveliotou, C., \& Patel, 
S.~K.\ 2001, \apjl, 563, L49 

\bibitem[Israel, Mereghetti, \& Stella(1994)]{Israel94} Israel, G.~L.,
Mereghetti, S., \& Stella, L.\ 1994, \apjl, 433, L25

\bibitem[Israel et al.(1999)]{Israel99} Israel, G.~L.~et al.\ 1999,
\aap, 346, 929

\bibitem[Juett et al.(2002)]{Juett02} Juett, A.~M., Marshall, H.~L., 
Chakrabarty, D., \& Schulz, N.~S.\ 2002, \apjl, 568, L31 

\bibitem[Kern \& Martin (2001)]{Kern01} Kern, B. \& Martin, C.\ 2001,
IAU Circular No. 7769

\bibitem[{{Markevitch} \& {Vikhlinin}(2001)}]{markevitch2001}
{Markevitch}, M. \& {Vikhlinin}, A. 2001, in Submitted to \apj,
  \\astro-ph/010593

\bibitem[{{Mereghetti}(1999)}]{mereghetti1999}
{Mereghetti}, S. 1999, in {\it The NS-BH Connection}, astro--ph/9912207

\bibitem[{{Mereghetti} {et~al.}(1998){Mereghetti}, {Israel}, \&
  {Stella}}]{mereghetti1998}
{Mereghetti}, S., {Israel}, G.~L., \& {Stella}, L. 1998, \mnras, 296,
689

\bibitem[Mereghetti \& Stella(1995)]{Mereghetti95} Mereghetti, S.~\&
Stella, L.\ 1995, \apjl, 442, L17

\bibitem[Morrison \& McCammon(1983)]{Morrison83} Morrison, R.~\& 
McCammon, D.\ 1983, \apj, 270, 119 

\bibitem[Paul, Kawasaki, Dotani, \& Nagase(2000)]{Paul00} Paul, B.,
Kawasaki, M., Dotani, T., \& Nagase, F.\ 2000, \apj, 537, 319

\bibitem[van Paradijs, Taam, \& van den Heuvel(1995)]{vanParadijs95} van
Paradijs, J., Taam, R.~E., \& van den Heuvel, E.~P.~J.\ 1995, \aap, 299,
L41

\bibitem[White et al.(1987)]{White87} White, N.~E., Mason, K.~O.,
Giommi, P., Angelini, L., Pooley, G., Branduardi-Raymont, G., Murdin,
P.~G., \& Wall, J.~V.\ 1987, \mnras, 226, 645

\bibitem[White et al.(1996)]{White96} White, N.~E., Angelini, L.,
Ebisawa, K., Tanaka, Y., \& Ghosh, P.\ 1996, \apjl, 463, L83

\bibitem[Wilson et al.(1999)]{Wilson99} Wilson, C.~A., Dieters, S.,
Finger, M.~H., Scott, D.~M., \& van Paradijs, J.\ 1999, \apj, 513, 464

\bibitem[{{Wise} {et~al.}(1997){Wise}, {Huenemoerder}, \&
{Davis}}]{wise2000}
{Wise}, M.~W., {Huenemoerder}, D.~P., \& {Davis}, J.~E. 1997, in ASP
Conf. Ser.
  125: Astronomical Data Analysis Software and Systems VI, Vol.~6, 477

\end{thebibliography}
\end{document}